# A Computational Study of Yttria-Stabilized Zirconia: I. Using Crystal Chemistry to Search for the Ground State on a Glassy Energy Landscape

Yanhao Dong[1], Liang Qi[2], Ju Li[3] and I-Wei Chen[1*]

[1]*Department of Materials Science and Engineering, University of Pennsylvania, Philadelphia, PA 19104, USA*

[2]*Department of Materials Science and Engineering, University of Michigan, Ann Arbor, MI 48109, USA*

[3]*Department of Nuclear Science and Engineering and Department of Materials Science and Engineering, Massachusetts Institute of Technology, Cambridge, MA 02139, USA*

**Abstract**

Yttria-stabilized zirconia (YSZ), a $ZrO_2$-$Y_2O_3$ solid solution that contains a large population of oxygen vacancies, is widely used in energy and industrial applications. Past computational studies correctly predicted the anion diffusivity but not the cation diffusivity, which is important for material processing and stability. One of the challenges lies in identifying a plausible configuration akin to the ground state in a glassy landscape. This is unlikely to come from random sampling of even a very large sample space, but the odds are much improved by incorporating packing preferences revealed by a modest sized configurational library established from empirical potential calculations. *Ab initio* calculations corroborated these preferences, which prove remarkably robust extending to the fifth cation-oxygen shell about 8 Å away.

[*]Corresponding Author
*E-mail address*: iweichen@seas.upenn.edu (I-Wei Chen)

Yet because of frustration there are still rampant violations of packing preferences and charge neutrality in the ground state, and the approach toward it bears a close analogy to glass relaxations. Fast relaxations proceed by fast oxygen movement around cations, while slow relaxations require slow cation diffusion. The latter is necessarily cooperative because of strong coupling imposed by the long-range packing preferences.

**Keywords:** Yttria-stabilized zirconia; Atomistic simulations; Atomic structure; EXAFS; Point defects; Glass

## I. Introduction

Zirconia solid solution is a prominent family of functional and structural ceramics that encompass both a cubic and a tetragonal form with many divalent and trivalent cation solutes. They exhibit high oxygen-ion ($O^{2-}$) conductivity, and are thus widely used as oxygen sensors and fuel/electrolysis cells, for which cubic yttria-stabilized zirconia (YSZ) is the most popular. Naturally, there are numerous experimental [1-3] and theoretical studies [3-6] on $O^{2-}$ diffusion. In comparison, there have been relatively few studies on cation diffusion, which in YSZ at 1000°C is at least one-trillion-fold slower than anion diffusion [7]. Such a sluggish kinetics ensures the stability of solid oxide fuel/electrolysis cells during their service since microstructural changes require coordinated motions of both cations and anions. For the same reason, ceramic processing—involving sintering and grain growth, being transport-limited by the slowest species—is controlled by cation diffusion, as is ceramic degradation and deformation at elevated temperature via creep and creep cavitation. Therefore, understanding cation kinetics is very important. Unfortunately, the previous theoretical and computational studies repeatedly and massively underestimated zirconia's cation diffusivity: They produced an activation energy of >10 eV, [8-11] which is clearly out of the realm of possibility and, indeed, is twice the experimentally observed value (~5 eV, which is in accordance with common practice of sintering YSZ at 1,300-1,550 °C) [7, 12]. The present study is thus undertaken to close this knowledge gap.

One of the computational challenges of YSZ lies in its structural complexity. For a zirconia solid solution that contains 8 mol% $Y_2O_3$, or 8YSZ, there are already an astronomically large number of possible configurations in a 3×3×3 supercell: Placing 92 Zr/16 Y on 108 cation sublattice sites and 208 O/8 $V_O$ on 216 anion sublattice sites allows $C_{108}^{16} \times C_{216}^{8} = 5.55 \times 10^{42}$ permutations. Such complexity has forced the past computer simulations on cation diffusion to rely on empirical potentials instead of *ab initio* ones [8-11], which may have been the reason for the unsatisfactory outcome. The problem is further complicated by the large distortions at most, if not all, the lattice sites in zirconia. Indeed, cubic zirconia solid solutions do not manifest cubic symmetry in the local structure according to Raman spectra [13, 14] and Extended X-ray Absorption Fine Structure (EXAFS) [15-18]; only when probed by diffraction techniques on a longer length scale is the cubic symmetry manifest. Naturally, the lack of symmetry levies further burden on computation regardless the size of the supercell. Therefore, we set our first task to develop a facile methodology to identify the most plausible and representative configurations that may be further studied by *ab initio* calculations. This is the subject of the present paper, which lays the foundation for the diffusion calculations in the companion paper.

To perform this task, we shall employ packing rules to help cope with structural complexity. Here we are inspired by a recent theoretical work that used the so-called bond valence rule to study Pb(Zr$_{1-x}$Ti$_x$)O$_3$ [19], which is also a concentrated solid solution with large distortions [20, 21]. Historically, chemists have established packing rules (also called crystal chemistry rules) by observing structural preferences in simple molecules and perfect crystals. Their physical rationale is self-evident: The mere existence of such structures is a testament of the energetic benefit of the rules. However, because of frustration in a defect-rich concentrated solid solution like YSZ, all the rules cannot possibly be satisfied at every level of local structure. Nevertheless, we shall hypothesize that *the more the rules are followed, the more stable the structure*, and we shall test this hypothesis computationally. In zirconia solid solutions, coordination preferences according to dopant studies by EXAFS [15-18, 22-24] and computational simulations [5, 6] are: Fewer O$^{2-}$ (i.e., more oxygen vacancies, V$_O$) around Zr$^{4+}$ than around Y$^{3+}$ in the first nearest neighbor (1NN) anion-shell of cations, and vice versa in the second nearest neighbor (2NN) anion-shell of cations. These preferences have been intuitively rationalized in terms of the cation sizes and formal charges. ((a) According to Pauling's rule, which states that a higher ratio of cation radius to anion radius favors a higher coordination number, Y$^{3+}$ being larger than Zr$^{4+}$ prefers higher coordination and accepts $V_O$ as 1NN; (b) $Zr^{4+}$ being smaller than $Y^{3+}$ prefers lower coordination and accepts $V_O$ as 1NN; and (c) while not accepting $V_O$ as 1NN, two $Y^{3+}$ being one valence lower than $Zr^{4+}$ each prefer to share one $V_O$ as 2NN, thus maintain charge neutrality.) If we can computationally verify our hypothesis, then we may incorporate these preferences into a protocol to more efficiently search for the most plausible low-energy structure. Hopefully, such structure is very close to the *ground state*, thus suitable for cation-defect and cation-diffusion calculations in the companion paper [25].

The remainder of this paper is as follows. Section II describes the computational methods. Section III constructs a library of configurations to establish the packing rules. Section IV employs the packing preferences to seek the minimum-energy configuration. The ground state configuration is further analyzed using the bond valence concept in Section V. This is followed by discussion in Section VI and conclusions in Section VII. Below, we will primarily study 8YSZ, which is of practical interest with many electrochemical applications; pure $ZrO_2$ will also be studied to provide a comparative reference.

**II. Methodology**

2.1 Empirical potential

The interatomic potential developed by Schelling *et al* [26], known to correctly predict both the cubic-to-tetragonal transition in zirconia and yttria's effect on stabilizing the cubic polymorph, and to also reasonably describe $O^{2-}$ diffusivity which peaks at 8YSZ [6], was employed in this study. The potential considers ions interacting via the Buckingham potential and the Coulombic potential

$$E_{ij} = A_{ij} \exp(-\frac{r_{ij}}{\rho_{ij}}) - \frac{C_{ij}}{r_{ij}^6} + \frac{q_i q_j}{r_{ij}} \quad (1)$$

where ions $i$ and $j$ separated by $r_{ij}$ have formal charges $q_i$ and $q_j$, respectively, and their species-dependent parameters $A$, $\rho$ and $C$ are given in Ref. [26]. To correspond to the composition of 8YSZ exactly, the simulation used a 3×3×3 supercell that contains 92 Zr ions, 16 Y ions and 208 O ions, and under the periodic boundary condition it performed positional relaxation and energy minimization at 0 K at zero pressure using General Utility Lattice Program (GULP) [27]. Some calculations were also performed using larger supercells, but the comparison with *ab initio* calculations will be for 3×3×3 supercells only.

2.2 *Ab initio* calculations

For such calculations, we used the projector augmented-wave (PAW) method [28] and the Perdew-Burke-Ernzerhof (PBE) [29] generalized gradient approximation (GGA) implemented in the Vienna *ab initio* simulation package (VASP) [30]. The PAW potentials include the following electrons: $5s^2 4d^2$ for Zr, $4s^2 4p^6 5s^2 4d^1$ for Y and $2s^2 2p^4$ for O. We chose a plane-wave cutoff energy of 500 eV to reach a convergence criterion of 1 meV for the total energy and sampled the Brillouin zone using the Monhorst-Pack scheme with a 2×2×2 $k$-point mesh. To validate the pseudopotentials and our method, calculations were first performed for 108 Zr and 216 O to obtain the three $ZrO_2$ polymorphs: cubic, tetragonal and monoclinic ones. As shown in **Table I**, the calculated lattice parameters and 0K transition enthalpies are in good agreement with the computed results [31] and the experimental data [32, 33] reported in the literature.

**Table I** *Ab initio* calculated lattice parameters and transformation enthalpies at 0K for cubic, tetragonal, and monoclinic $ZrO_2$ against the experimental and simulation data in the literature.

|  | This work | Previous work | |
| --- | --- | --- | --- |
|  |  | Experimental [32, 33] | Simulation [31] |
| Cubic (c) | | | |
| Volume ($Å^3$) | 32.97 | 32.97 | 32.97 |
| a (Å) | 5.09 | 5.09 | 5.09 |
| Tetragonal (t) | | | |
| Volume ($Å^3$) | 33.49 | 33.04 | 34.55 |
| a (Å) | 3.603 | 3.571 | 3.628 |
| c (Å) | 5.160 | 5.182 | 5.250 |
| Monoclinic (m) | | | |
| Volume ($Å^3$) | 35.17 | 35.22 | 36.05 |
| a (Å) | 5.136 | 5.150 | 5.192 |
| b (Å) | 5.268 | 5.212 | 5.265 |
| c (Å) | 5.267 | 5.315 | 5.358 |
| β (º) | 99.25 | 99.23 | 99.81 |

| | Transformation enthalpy | | |
|---|---|---|---|
| t-c (eV) | -0.06 | -0.06 | -0.07 |
| m-c (eV) | -0.18 | -0.12 | -0.17 |

**III. Supercell energy and its correlation with coordination numbers**

3.1 A library of randomly sampled configurations: empirical potential calculations

To have a statistical glimpse of the energetics of the $5.55\times10^{42}$ configurations in a 3×3×3 supercell, we used empirical-potential calculations to relax 100,000 randomly generated configurations. The resulting energies shown in **Fig. 1** follow a Gaussian distribution spanning 15 eV. Setting the lowest energy as 0 eV, we find the average at 5.6 eV with a standard deviation of 1.4 eV. The distribution is robust: When we divide the 100,000 configurations into 10 subsets, we find they all have the same average and standard deviation as above.

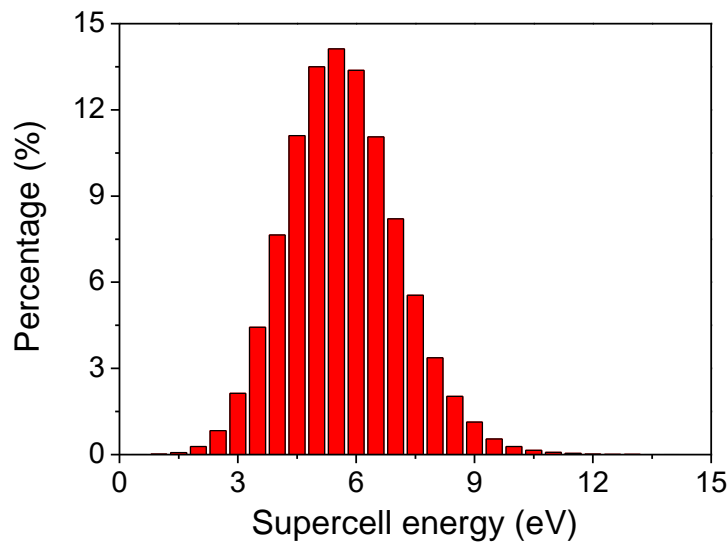

**Figure 1** Energy distribution of 100,000 randomly generated configurations, relaxed

by empirical potential calculations.

Since the probability of finding a higher energy configuration is handicapped by the Boltzmann factor, a 5.60 eV higher energy represents a $7.73\times10^{-15}$ ($\exp(-\Delta E/k_B T)$) times lower probability even at 2000 K. So most configurations in **Fig. 1** except those at the very low-energy end are unlikely to exist in reality. Indeed, as will be demonstrated later, the ground state of the $5.55\times10^{42}$ configurations must have a very negative energy in the scale of **Fig. 1**. Therefore, the ground state is *not* a typical configuration, and it *cannot* be efficiently searched by random sampling. On the other hand, although the above configurational library is useless for finding the ground state, one can mine it to uncover the preferred crystal chemical arrangements, as shown below.

3.2 Crystal chemistry preferences

We now analyze the library to find the correlation between energies and local environments, thereby establish the packing rules. Despite the highly defective structure and large lattice distortions, we found the average cation-centered cation-oxygen radial distribution functions (RDF) $g(r)$ are quite robust. The RDFs shown in **Fig. 2** are averages from 100 randomly selected configuration, but they are also indistinguishably reproduced by other average RDFs using other randomly selected configurations. Here, $g(r)$

$$g_{\text{Zr-O}}(r) = \frac{Vol}{N_\text{O}} \frac{\frac{1}{N_{\text{Zr}}} \sum_{i=1}^{N_{\text{Zr}}} n_{\text{O},i}}{4\pi r^2 \Delta r} \qquad (1)$$

$$g_{\text{Y-O}}(r) = \frac{Vol}{N_\text{O}} \frac{\frac{1}{N_{\text{Y}}} \sum_{i=1}^{N_{\text{Y}}} n_{\text{O},i}}{4\pi r^2 \Delta r} \qquad (2)$$

is obtained by (a) counting the number of O neighbors ($n_{\text{O},i}$) situated between distances $r$ and $r+\Delta r$ from the $i^{\text{th}}$ Zr or Y cation among $N_{\text{Zr}}$=92 Zr cations or $N_{\text{Y}}$=16 Y cations, and (b) constructing a dimensionless average of the distribution over the supercell by normalizing the sum of $n_{\text{O},i}$ by the total oxygen number ($N_{\text{O}}$=208) and the supercell volume (*Vol*). Despite some peak broadening brought upon by disorder and distortion in the local structures, the RDFs have clearly defined, well separated peaks for at least up to the 5$^{\text{th}}$ oxygen shells. Note that the smaller ionic radius of $Zr^{4+}$ is reflected in the first peak: The Zr-O distance is shorter than the corresponding Y-O distance. But in the second shell there is no apparent difference between the Zr-O and Y-O distance. Therefore, the ion-size effect in YSZ appears to be rather short-ranged. Meanwhile, the cation-cation RDFs (Zr-Zr, Zr-Y, Y-Y, Y-Zr), also plotted in **Fig. 2**, all have the same peaks at the same distance starting from the 1NN cation-cation at 3.6 Å.

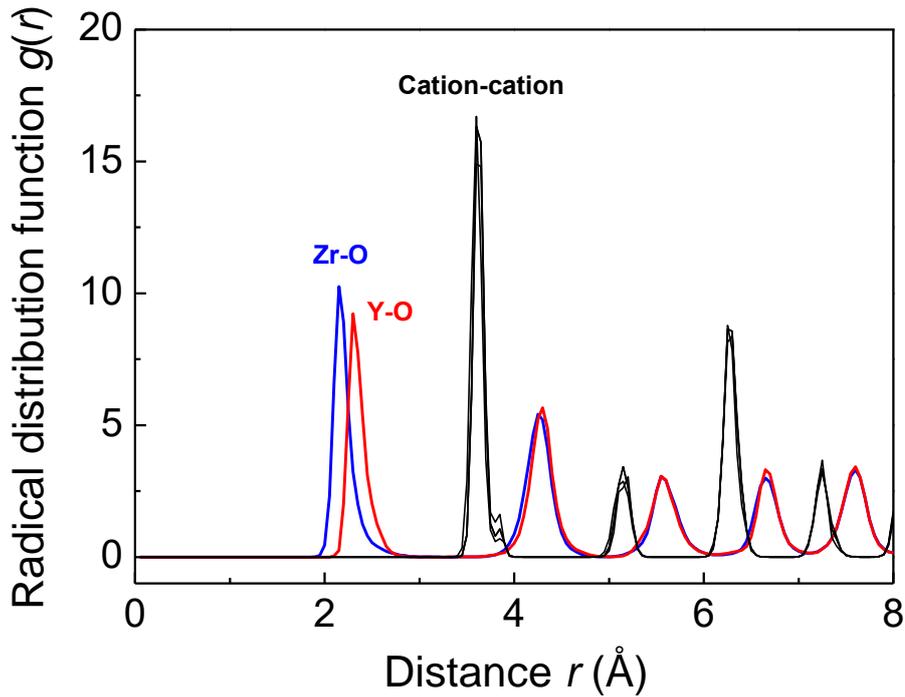

**Figure 2** Radial distribution functions centered around Zr (blue) and Y (red) of 100 randomly selected configurations according to empirical potential calculations. The first peak is due to cation-oxygen coordination. Also included are cation-cation radial distribution functions (black), which include four separate ones for Zr-Zr, Zr-Y, Y-Zr and Y-Y but indistinguishable from each other.

Our result on cation-cation RDFs is consistent with the EXAFS finding of Li *et al.* [16, 17], that the cation-cation distance is nearly unchanged in all cubic and tetragonal zirconia regardless of the concentration, size and charge of dopants, including $Ga^{3+}$, $Fe^{3+}$, $Y^{3+}$, $Gd^{3+}$, $Ge^{4+}$, $Ce^{4+}$, and $Nb^{5+}$ (codoped with $Y^{3+}$). This was explained by (a) the $r^{-2}$ decay of elastic distortion (in an ideal fluorite structure the closest cation-cation distance is $(8/3)^{1/2}$ times the closest cation-O distance), and (b) the relative softness of oxygens compared to cations. This explanation also accounts for the rapid washout of the size effect on the Zr/Y-O RDF beyond the 1NN. Because of this, and because the 2NN Zr/Y-O peak falls between the 1NN and 2NN cation-cation peaks, it is impossible to use EXAFS to definitely identify the 2NN Zr/Y-O, contrary to the conclusions of many who incorrectly cited Reference [16]. (As pointed out by Li *et al.*, the cation EXAFS that benefits from much stronger photoelectron scattering as well as multiple scattering can completely mask the Zr/Y-O EXAFS beyond the 1NN, and this is especially true in cubic zirconia that is more distorted than tetragonal zirconia.) Therefore, we will not attempt to directly compare our RDFs (both Zr/Y-O and Zr/Y-(Zr/Y)) with EXAFS data further.

However, our computational study can reveal much more information than RDFs because it has access to all the atomic positions, which allows us to use the following method to calculate the average coordination numbers (within the supercell) of every shell. Irrespective of the local structure distortions that can be quite severe because of the copious $V_O$ population, this method is exact since it is based on the fact that each $O^{2-}$ must be fully bonded—having 4 cations as the 1NN and 12 cations as the 2NN, etc.—in a supercell that has no cation vacancy. Counting these Zr or Y neighbors around each O and summing them up over all 208 oxygens in the supercell, one obtains the total numbers of Zr-O and Y-O pairs as the 1NN, 2NN, etc. Dividing these pair numbers by the cation numbers (92 Zr and 16 Y) thus yields the average Zr-O and Y-O pair numbers around each Zr/Y, which is precisely the average O-coordination number of Zr/Y.

This method provides the detailed statistics in **Fig. 3** of average coordination number that manifests a strong correlation with the supercell energy. The distributions may be compared with the numbers of anion sites around a cation site in a perfect fluorite structure (space group $F m \bar{3} m$), which is the reference structure of cubic $ZrO_2$: There are 8 1NN, 24 2NN, 24 3NN, 32 4NN, and 48 5NN. Around Zr, a lower O coordination number is correlated to a lower supercell energy in the 1NN, 3NN and 5NN, and a higher supercell energy in the 2NN and 4NN. To quantify the trends, we define a correlation coefficient $\rho(x,y)$

$$\rho(x, y) = \frac{E[(x - \bar{x})(y - \bar{y})]}{\sigma_x \sigma_y} \quad (3)$$

where $\bar{x}$ and $\bar{y}$ are the averages, $\sigma_x$ and $\sigma_y$ are the standard deviations, and $E$ computes the expectation value of the variable for the distribution. Note that the sign of the correlation coefficient for the coordination number and supercell energy (listed in **Fig. 3**) alternates from shell to shell, and its absolute value peaks at the 2NN of Zr. Note also that an opposite set of correlations exists around Y as shown in the insets of **Fig. 3.** The latter results can be easily understood because in each shell the total number of Y's O neighbors plus the total number of Zr's O neighbors must equal to the total number of O's cation neighbors, which is 208 times 4 for the 1NN, 208 times 12 for the 2NN, etc. As the total number of O (or Ov) in each shell for all the cations must conserve, any deficit in the O (or Ov) concentration around Zr must be made up by an excess in the same concentration around Y within the same shell, and vice versa. This result also agrees with the calculated binding energies for Y-$V_O$ pair, which peaks at the same distance as that of 2NN [5, 6]. Therefore, the overall trend is self-consistent.

To understand why the 2NN correlation is the strongest, we will use the following simple argument. There are two effects in YSZ: size and charge. For 2NN, the two effects reinforce each other; for 1NN, they counter each other, and the size effect has an upper hand. This is most easily seen for $Y^{3+}$, which differing from the host $Zr^{4+}$ in charge and in size is the origin of packing preferences. From the charge consideration (Coulombic interactions), $Y^{3+}$ of a lower valence prefers to be less surrounded by $O^{2-}$, hence more $V_O$ around, as both 1NN and 2NN. From the size consideration (Pauling's rule, which comes from short-range repulsion and is distinct from elastic interaction), $Y^{3+}$ being larger than $Zr^{4+}$ prefers a higher coordination number, thus no $V_O$ as 1NN. These two effects counter each other in 1NN, but in 2NN they reinforce each other both preferring Y-$V_O$ 2NN pairing. This is why the correlation for 2NN is stronger than that for 1NN.

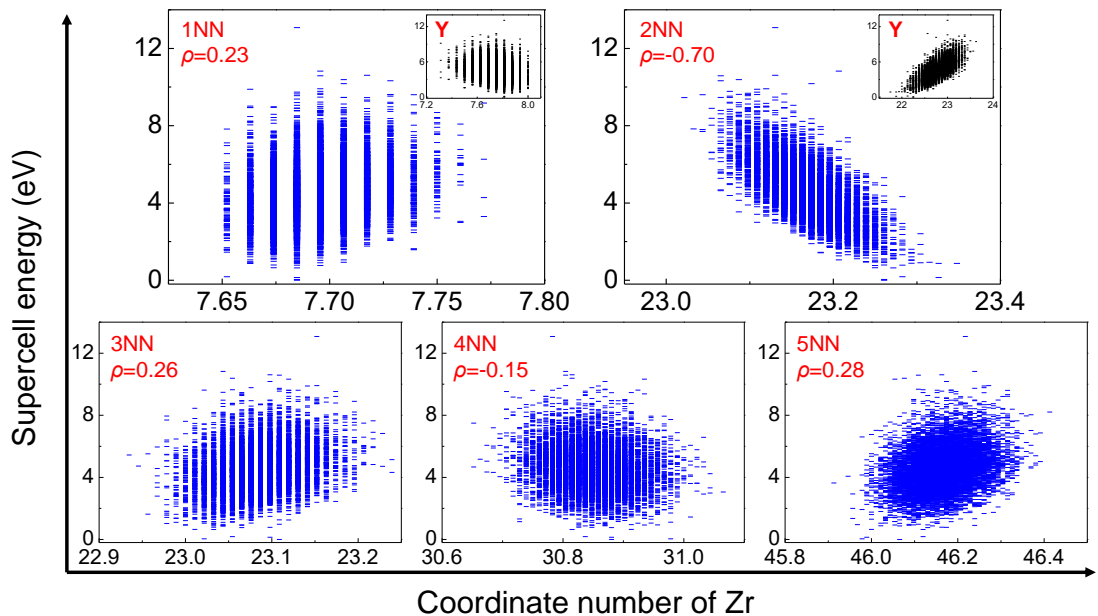

**Figure 3** Average number of oxygens in 1NN to 5NN oxygen shells around Zr, with each data point (shown as horizontal bar) corresponding to one configuration according to empirical potential calculation. 100,000 configurations in total. Inset: Corresponding data around Y.

To quantify the packing preferences, we calculated the average $V_O$ concentration around a center cation (Zr or Y) for the 100 lowest energy configurations (corresponding to the low-energy tail in **Fig. 1**). The results shown in **Fig. 4a** follow a distinct oscillatory pattern from shell to shell around the average $V_O$ concentration—8 $V_O$ out of 216 anion sites or 3.076%. Again, the oscillations are of opposite sense for Zr and Y because the total number of $V_O$ must conserve in each shell. Likewise, we calculated the average Y concentration around an O, shown in **Fig. 4b**, which is also oscillatory around the average Y concentration, 16 Y out of 108 cation sites or 14.81%. (Not shown is the average Zr concentration, being 100% minus the average Y concentration, which follows an opposite oscillatory pattern.) Remarkably, these concentration oscillations are surprisingly robust and long-ranged, persisting to the $5^{th}$ anion-shell of Zr and Y and to the $5^{th}$ cation-shell of O, or 8 Å according to **Fig. 2**, which is already half the size of the supercell (15 Å). *That is, the concentration oscillations can outlast the washout of atomic strain undulations in **Fig. 2** and are felt throughout the supercell.* Specifically, $V_O$ prefers to be the 1NN, 3NN and 5NN of Zr, and 2NN and 4NN of Y; correspondingly, Y prefers to be the 1NN, 3NN and 5NN, but not the 2NN and 4NN of O. In agreement with the finding in **Fig. 3**, the 2NN $V_O$-Y correlation is the strongest. To summarize these results, we list in **Table II** some statistical parameters for the data in **Fig. 3-4**. To provide some indications that these results are not due to the small cell size, we also performed calculations using 4×4×4 supercells containing 218 Zr, 38Y and 493 O and obtained similar results. (See **Table II**.) For both sets of supercells, we found the highest values of $\frac{1}{N_C}\frac{dE}{dCN_{Zr}}$, $\frac{N_0}{N_C}\frac{dE}{dCN_{Zr}}$, $\rho_{Zr}$, $\Delta[V_O]_Y$, and $\frac{\Delta[V_O]_Y}{[V_O]_Y}$ (see definitions in **Table II**) at the 2NN correlations.

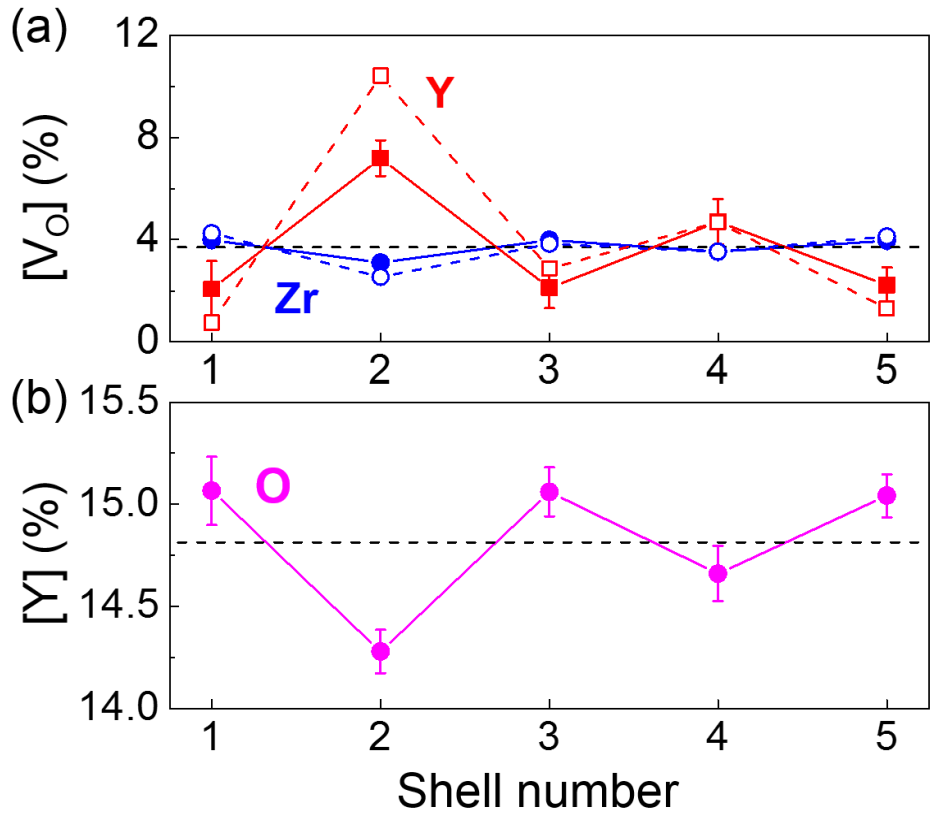

**Figure 4** (a) Average oxygen vacancy concentration around a center Zr (blue) or Y (red) oscillates from shell to shell, and (b) average Y concentration around a center O oscillates from shell to shell. (The Zr concentration is 100%-[Y] concentration.) These averages (solid symbols) are from 100 lowest energy configurations in **Fig. 1**.

Also shown in (a) in open symbols are *ab initio* calculated data (see **Table III** for details) for the *de-facto* ground state (the red cross in **Fig. 5**).

**Table II** Summary of number of oxygens in the 1st to 5th coordination shell of Zr in the supercell; data for 3×3×3 supercell averaged over 100,000 configurations as shown in Fig. 3; data for 4×4×4 supercell averaged over 1,000 random configurations. $N_0$: reference data for ZrO$_2$ in fluorite structure. $\frac{1}{N_C}\frac{dE}{dCN_{Zr}}$: slopes in Fig. 3 normalized by the total number of cation $N_C$=108 in the 3×3×3 supercell; or normalized slopes calculated for 4×4×4 supercell in the same way with $N_C$=256. $\frac{N_0}{N_C}\frac{dE}{dCN_{Zr}}$: same as the former one, normalized by $N_0$. $\rho_{Zr}$: the correlation factor. Summary of average concentration of oxygen [V$_O$] in the 1st to 5th coordination shell of Y in the supercell; data for 3×3×3 supercell averaged over 100 lowest energy configurations as shown in Fig. 4; data for 4×4×4 supercell averaged over 10 lowest energy configurations from 1,000 random configurations. $\Delta[V_O]_Y$: deviations of [V$_O$] from the mean. $\frac{\Delta[V_O]_Y}{[V_O]_Y}$: same as previous, normalized by the mean. $\sigma_Y$: standard deviation.

| | 3×3×3 supercell | | | | | | 4×4×4 supercell | | | | | |
|---|---|---|---|---|---|---|---|---|---|---|---|---|
| $N_0$ | $\frac{1}{N_C}\frac{dE}{dCN_{Zr}}$ (eV) | $\frac{N_0}{N_C}\frac{dE}{dCN_{Zr}}$ (eV) | $\rho_{Zr}$ | $\Delta[V_O]_Y$ | $\frac{\Delta[V_O]_Y}{[V_O]_Y}$ | $\sigma_Y$ | $\frac{1}{N_C}\frac{dE}{dCN_{Zr}}$ (eV) | $\frac{N_0}{N_C}\frac{dE}{dCN_{Zr}}$ (eV) | $\rho_{Zr}$ | $\Delta[V_O]_Y$ | $\frac{\Delta[V_O]_Y}{[V_O]_Y}$ | $\sigma_Y$ |
| 1NN | 8 | 0.156 | 1.25 | 0.23 | -0.016 | -0.44 | 0.011 | 0.222 | 1.78 | 0.33 | -0.018 | -0.48 | 0.007 |
| 2NN | 24 | -0.237 | -5.69 | -0.70 | 0.035 | 0.94 | 0.007 | -0.205 | -4.91 | -0.62 | 0.027 | 0.73 | 0.006 |
| 3NN | 24 | 0.086 | 2.06 | 0.26 | -0.016 | -0.43 | 0.008 | 0.062 | 1.48 | 0.17 | -0.008 | -0.21 | 0.007 |
| 4NN | 32 | -0.037 | -1.19 | -0.15 | 0.010 | 0.27 | 0.009 | -0.016 | -0.50 | -0.06 | 0.008 | 0.21 | 0.007 |
| 5NN | 48 | 0.058 | 2.78 | 0.28 | -0.015 | -0.40 | 0.007 | 0.043 | 2.07 | 0.21 | -0.010 | -0.26 | 0.005 |

3.3 Packing preferences confirmed by *ab initio* calculations

The results in **Fig. 3-4** can be naively interpreted as crystal chemistry preferences for atomic bonding. Below we will verify them by *ab initial* calculations. To compare with empirical-potential calculations, we first selected 11 representative configurations from the 100,000 configurations in **Fig. 1**, ones that have their supercell energies differing in 1.0 eV increments, so that they span the whole energy range in **Fig. 1**, from 0 to 10.0 eV. For each of them, we let its already relaxed configurations to further relax in the *ab initio* calculation to arrive at the final state. As shown in **Fig. 5**, the supercell energies obtained by the two sets of calculations track each other reasonably well; here for ease of comparison we set the lowest energy obtained by the 11 *ab initio* calculations to be 0 on the *ab initio* scale, and likewise set the lowest energy obtained by the 11 empirical potential calculations to be 0 on the empirical potential scale. Next, we analyzed the local environments of the 11 *ab initio* calculated configurations in the same way as in **Fig. 3**, and found them (shown in **Fig. 6**) to follow the same crystal chemistry preferences including the long-range features. Remarkably, although we expected their statistics to be poorer because of the much smaller sample size (11 vs. 100,000), we actually found— except in one case (3NN of Zr)—their $|\rho|$ to be higher than its counterpart in **Fig. 3**, probably because the supercell energies in this small sample space are on average lower. These results affirm: (a) $V_O$ prefers to be the 1NN, 3NN and 5NN of Zr, and the 2NN and 4NN of Y; (b) the 2NN $V_O$-Y has the strongest correlation, and (c) these are all the long-ranged correlations persisting to the 5$^{th}$ shell. A visual demonstration of these preferences is again presented in **Fig. 4** in which the *ab initio* calculated structure of the *de-facto* ground state (the red cross in **Fig. 5**, more on it later) exhibits the same oscillatory feature as the empirical-potential-calculated structures of 100 lowest energy states.

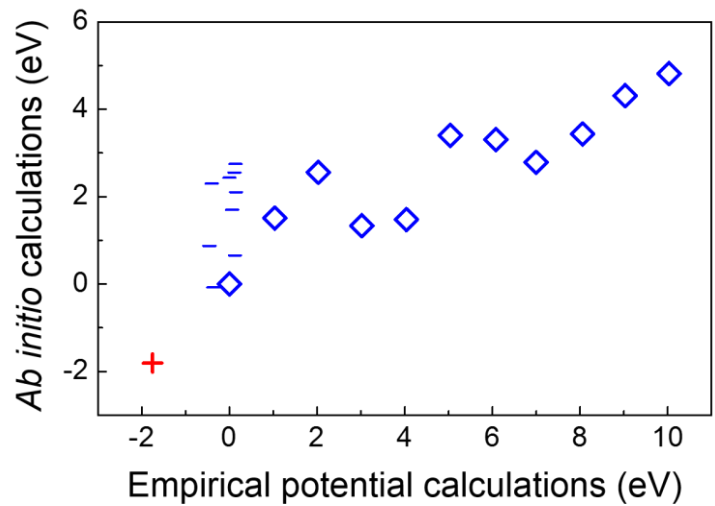

**Figure 5** Supercell energy according to empirical potential and *ab initio* calculations. Each data point corresponds to one starting configuration, and in each calculation method the configuration that has the lowest energy sets the zero-energy for said calculation method. Also included are the 10 lowest energy configurations in **Fig. 1** (blue dashes, one overlapping the lowest diamond) and the *de-facto* "ground" state (red cross).

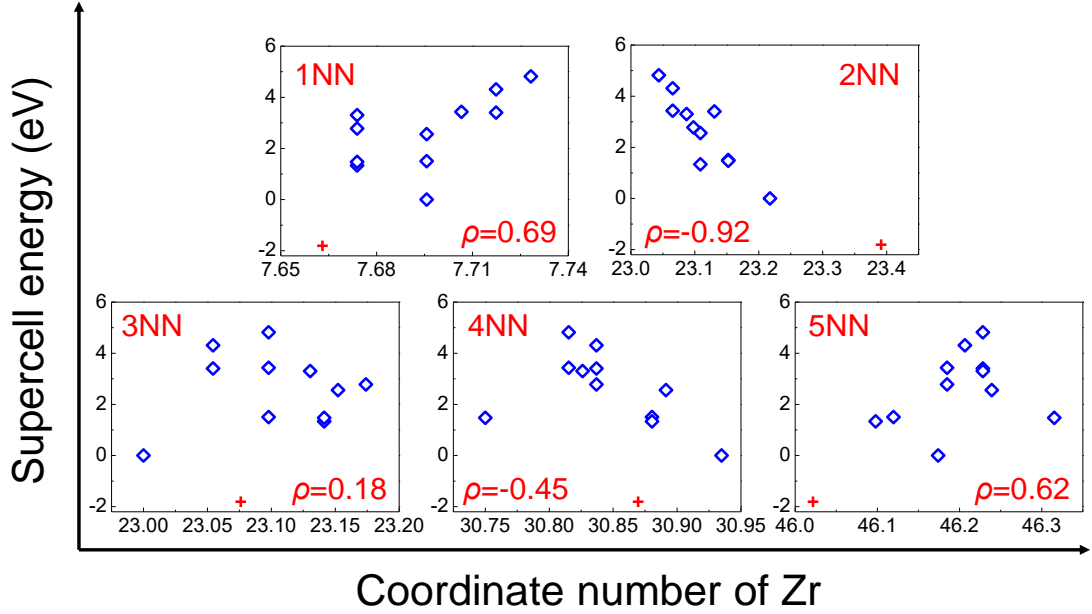

**Figure 6** Average number of oxygens in 1NN to 5NN oxygen shells around Zr, with each data point corresponding to one configuration according to *ab initio* calculation. Blue diamonds are the same configurations in Fig. 5, and red cross is the "ground" state configuration.

## IV. Configurations akin to the ground state

4.1 Seeking minimum energy configurations using packing preferences

The remarkably robust and long-ranged packing preferences revealed by the above correlation study support our hypothesis: They have a decisive influence on the supercell energy. Can they help find the lowest energy configuration more efficiently?

To answer this question, we first employed one such preference—$V_O$'s preference to be the 2NN of Y—to manually build the following sets of configurations. They have ion arrangements that satisfy this preference to various extents, quantified by the number $n_{\text{Y-(VO-2NN)}}$ of Y-($V_O$-2NN) pairs in the supercell, and for each $n_{\text{Y-(VO-2NN)}}$ ($n_{\text{Y-(VO-2NN)}}$ from 24 to 52, as it turns out) we obtained 100 configurations using the following procedure. (a) Start with a random distribution of Zr and Y on the cation sublattice. (b) Rank the anion sublattice sites by the numbers of their Y-2NN. (c) Place $V_O$ at the anion sites starting with the ones with the largest number of Y-2NN. (d) Continue with (c) at the remaining sites, selected in the order of decreasing number of Y-2NN, until the desired $n_{\text{Y-(VO-2NN)}}$ is obtained. (e) If (d) fails to obtain such preset $n_{\text{Y-(VO-2NN)}}$, then repeat (a-d). In this way, we obtained sets with $n_{\text{Y-(VO-2NN)}}$ as large as 52. (Higher numbers are hindered by frustrations under the compositional constraint, since the placement of $V_O$ depends not only on Y's preference but also Zr's.) We can gain some perspective on $n_{\text{Y-(VO-2NN)}}$ by recalling each $V_O$ has 12 cation-2NN, each supercell has 8 $V_O$, and 14.81% of cation sites are occupied by Y. Thus, the theoretical maximal for $n_{\text{Y-(VO-2NN)}}$ should be 96 and the average $n_{\text{Y-(VO-2NN)}}$ in random sampling should be 14.2. However, interactions between supercells imposed by the periodic boundary condition make it difficult to obtain too high an $n_{\text{Y-(VO-2NN)}}$, which by dictating the 2NN preference reflects a longer range correlation that may be frustrated the intercell interactions. Indeed, many high $n_{\text{Y-(VO-2NN)}}$ configurations turn out to be unstable as described below.

To assess the stability of these configurations, we performed empirical-potential relaxations to obtain their energy distributions in **Fig. 7a**. The distributions show a large spread especially in the high $n_{\text{Y-(VO-2NN)}}$ set. This is because some of their configurations are unstable; during relaxation, some O will spontaneously relocate so the configuration acquires a different $n_{\text{Y-(VO-2NN)}}$. One such example is shown in **Fig.**

**7b** for the set that initially has $n_{\text{Y-(VO-2NN)}}$=52; after relaxation, no configuration has $n_{\text{Y-(VO-2NN)}}$=52; instead, $n_{\text{Y-(VO-2NN)}}$ varies from 24 to 49. Despite the relocation, it is clear that the sets with a higher $n_{\text{Y-(VO-2NN)}}$ still tend to have a lower energy. Moreover, within each set, the lowest energy configurations after allowing for relocation tend to achieve the highest post-relaxation $n_{\text{Y-(VO-2NN)}}$ (for example, in **Fig. 7b**, the two lowest energy configurations are at $n_{\text{Y-(VO-2NN)}}$=44 and 47.) In **Fig. 7a**, we have marked the lowest energy in each set by a red box to make it clear that the lowest set-energy generally decreases with the set $n_{\text{Y-(VO-2NN)}}$. The lowest energy is located within set $n_{\text{Y-(VO-2NN)}}$=52 with an energy –0.8 eV, i.e., it is 0.8 eV lower than the lowest energy in **Fig. 1**. This is remarkable: The $n_{\text{Y-(VO-2NN)}}$=52 set samples only 100 configurations, while **Fig. 1** samples 100,000 configurations, yet it is the 100 configurations set up with the aid of one packing preference that come closer to the ground state. If we judge the search efficiency by the inverse of the sample size traversed, then the use of one packing preference has already boosted the efficiency by 1,000 times.

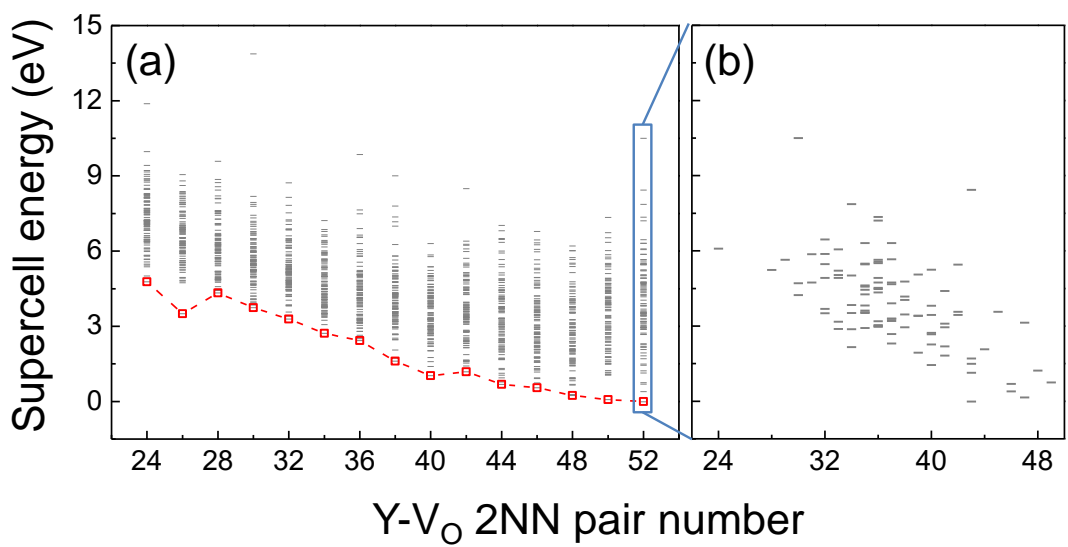

**Figure 7** (a) Supercell energies of manually built configurations according to empirical potential calculations with Y-$V_O$ 2NN pair numbers ($n_{\text{Y-(VO-2NN)}}$) preset from 24 to 52. Each column contains 100 configurations and the red boxes are the lowest energy configurations. (b) After relaxation according to empirical potential calculations, the actual $n_{\text{Y-(VO-2NN)}}$ differs from the starting $n_{\text{Y-(VO-2NN)}}$=52. The low energy configurations are ones that maintain a high $n_{\text{Y-(VO-2NN)}}$.

We next investigated whether another preference—$V_O$'s preference to be the 1NN of Zr—can help find a lower energy configuration. A similar search maximizing the number $n_{\text{Zr-(VO-1NN)}}$ of Zr-($V_O$-1NN) pairs returned sets (each having 100 configurations) with up to $n_{\text{Zr-(VO-1NN)}}$=32, with the lowest energy coming out from set $n_{\text{Zr-(VO-1NN)}}$=32 at 1.8 eV. (Since each $V_O$ has 4 cation-1NN, each supercell has 8 $V_O$, and 85.19% of cation sites are occupied by Zr, the theoretical maximum for $n_{\text{Zr-(VO-1NN)}}$ is 32 and the average $n_{\text{Zr-(VO-1NN)}}$ in random sampling is 27.5. Unlike the case of $n_{\text{Y-(VO-2NN)}}$, here we were able to obtain the theoretical maximum for the 1NN preference because it is a shorter range correlation.) Therefore, while maximizing the number of pairs was again beneficial in this search, the lowest energy configuration obtained in the search was not as low in energy as that in **Fig. 7a-b**. In this sense, this packing preference is less effective than the previous one.

Finally, we combined the two preferences by assigning a weight $w$ to favor having Y-($V_O$-2NN) of Y, and a complementary weight 1-$w$ to favor having Zr-($V_O$-1NN). This search produced the best results as shown in **Fig. 8**. (See a more detailed description of the procedure in the figure caption.) In this figure, six sets of configurations of 1,000 each were generated for $w$ ranging from 0 to 1, and the lowest energy configuration is from the $w$=0.8 set at –1.8 eV, i.e., its energy is 1.8 eV lower than the lowest energy in **Fig. 1**. This search partially benefited from a larger sample size than that in **Fig. 7a-b**, since it had 1,000 configurations instead of 100. But in the $w$=1 set its lowest energy is –1.5 eV, which is lower than that in **Fig. 7a-b** but not as low as in the $w$=0.8 set. Therefore, there is indeed an advantage of combining two packing preferences.

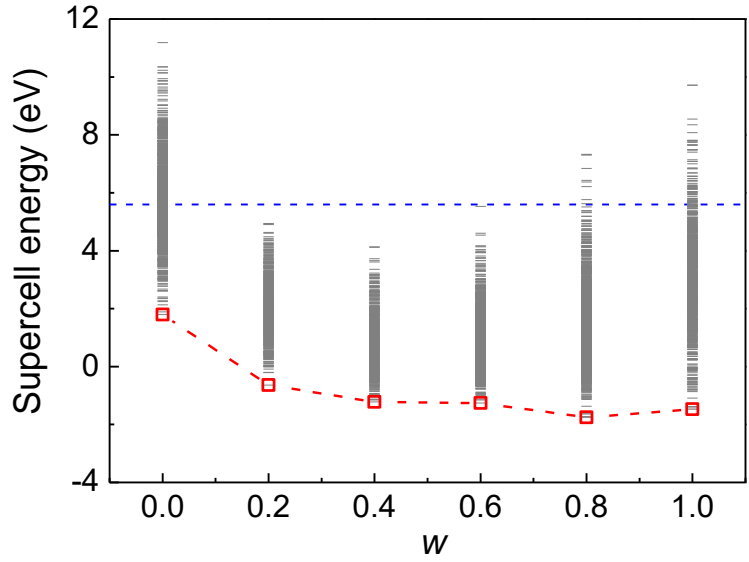

**Figure 8** Supercell energies of manually built configurations according to empirical potential calculations with preset weight $w$ from 0 to 1 to favor the highest preset $n_{Zr-(V_O-1NN)}$ 1NN vs. highest preset $n_{Y-(V_O-2NN)}$. Each column contains 1,000 configurations and the red boxes are the lowest energy configurations. The sampling procedure is as following: (a) Start with a random distribution of Zr and Y on the cation sublattice. (b) Rank the anion sublattice sites by the numbers of $\{(1-w)\times$(pair number of Zr-($V_O$-1NN)) + $w\times$(pair number of Y-($V_O$-2NN))$\}$. (c) Place $V_O$ at the site(s) starting with the highest numbers calculated in (b). (d) Continue (c) at the remaining sites, until the preset number of $\{(1-w)\times$(pair number of Zr-($V_O$-1NN)) + $w\times$(pair number of Y-($V_O$-2NN))$\}$ is obtained. (e) If (d) fails and such preset number cannot be obtained, then repeat (a-d). The preset numbers for each $w$ is chosen such that one configuration can be found after trial about 100,000 random cation distributions.

4.2 The "ground" state

To check whether the above lowest-energy configuration selected with the aid of packing preferences and empirical-potential calculations indeed has a lower energy than all the ones from random sampling, we further relaxed the configuration using *ab initio* calculation. We also studied the 10 lowest energy configurations in **Fig. 1** using *ab initio* calculation for further comparison. The results are plotted in **Fig. 5**: It shows that the former (red cross in **Fig. 5**) has the lowest energy, 1.8 eV less than the lowest of the latter 10 (blue dashes in **Fig. 5**). Note that in **Fig. 5** the ~5 eV span of the *ab initio* calculated energy corresponds to a 10 eV span of the empirical-potential-calculated energy, so we may equate the 1.8 eV decrement in the *ab initio* energy to a 3.6 eV decrement in the empirical potential energy. This is truly a huge advantage, and such low-energy configuration is unlikely to come from random sampling even if we enlarge the sample size (currently 100,000 in **Fig. 1**) by many orders of magnitude. Since this was the best configuration we obtained in our study, we will refer to it as the "ground" state from now on.

In this "ground" state, there are considerable variations in the local structure. For example, the number of O ranges from 6 to 8 in the 1NN around Zr and from 7 to 8 around Y; in the 2NN from 22 to 24 around Zr and from 19 to 23 around Y; in the 5NN from 43 to 48 around Zr and from 44 to 48 around Y. These variations exist because the compositional constraint and the periodic condition prevent the supercell from satisfying all the packing preferences everywhere, i.e., there is frustration. However, the O coordination numbers averaged over the supercell, listed in **Table III**, do have all the *right* features: The coordination number oscillates from shell to shell starting with $V_O$ favored in the 1NN of Zr, whereas the oscillation around Y is of the opposite sense, and in **Fig. 4**, the "ground" state structure exhibits the same oscillatory feature as empirical-potential-computed structures of 100 lowest energy states. A quantitative perspective is obtained by plotting these supercell-averaged data as red crosses in **Fig. 6**. For the 1NN, 2NN and 5NN, the ground state local environments follow the preferences significantly more than the 11 randomly selected configurations do. While we should caution that the statistics gleaned from one configuration in a 3×3×3 supercell is too sparse, we nonetheless see in these results that the much stricter adherence to the packing preferences is probably what makes the "ground" state the lowest energy state.

**Table III** Average coordination number from the $1^{st}$ to $5^{th}$ coordination shell of Zr and Y in a supercell in the "ground" state. Also shown are average bond valence, bond valence energy and electrostatic energy of the state. Configuration relaxed by *ab initio* calculations.

|  | Zr | Y |
|---|---|---|
| 1NN | 7.66 | 7.94 |
| 2NN | 23.39 | 21.50 |
| 3NN | 23.08 | 23.31 |
| 4NN | 30.87 | 30.50 |
| 5NN | 46.02 | 47.38 |
| Bond valence | 3.73 | 3.34 |
| Bond valence energy | | 9.27 |
| Electrostatic energy (eV) | | -7.01 |

4.3 Free energy

Metastable configurations do make a contribution to the free energy by way of entropic energy. To get a sense of the contribution of vibrational and configurational entropies, we calculated them for the 100,000 randomly generated configurations in **Fig. 1**. The vibrational entropy ($S$) at 1,000K (about 1/3 of the melting point) was obtained from the phonon spectra calculated in GULP using the same empirical potential. The resultant entropic energy ($-TS$) from the 100,000 random sampling follows a Gaussian distribution with a standard deviation of 0.25 eV and a mean of –130.7 eV, compared to the lowest energy of –11,400.9 eV. Importantly, it manifests no obvious correlation with the supercell energy over the entire energy range displayed in **Fig. 1**. To assess the contribution from the configurational entropy, we evaluated the ensemble average using the Boltzmann statistics in two cases. (a) Starting with the 0K energies for all the configurations in **Fig. 1**, we calculated the ensemble average of the entire distribution at 1,000K. (b) We repeated (a) but additionally included the vibrational entropy. Relative to the lowest energy in **Fig. 1**, set as 0 eV, the two ensemble averages are 0.08 eV for (a) and 0.11 eV for (b). Therefore, the Boltzmann statistics dictates that only states within about 0.1 eV above the lowest enthalpic-entropic energy are represented at 1,000K. Since vibrational entropy shows no obvious correlation throughout the distribution in **Fig. 1**, it cannot influence the state selection at finite temperature and all the configurations of relevance to the free energy calculation must still lie at the very left end of the distribution. Although the above calculations were made using empirical potential, we believe the conclusion should hold in general in view of the good correspondence in **Fig. 5** between empirical-potential calculations and *ab initio* calculations.

**V. Bond Valence Analysis**

To further check the "soundness" of the local atomic environments obtained by the *ab initio* calculations, we performed a bond valence analysis, which is an extension of Pauling's rules. The bond valence between cation Zr or Y (written below as Zr/Y) and its $i^{\text{th}}$ O-1NN is defined as

$$s_{\text{Zr/Y},i} = \exp(\frac{R_{\text{Zr/Y},0} - R_{\text{Zr/Y},i}}{B}) \quad (4)$$

where $R_{\text{Zr/Y},i}$ is the "observed" bond length between a cation and its $i^{\text{th}}$ O-1NN, $R_{\text{Zr/Y},0}$ is a tabulated parameter provided by Brown and Altermatt [34] that represents the "ideal" bond length of the "ideal" valence, and $B$ is an empirical constant typically set at 0.37 Å. Following Pauling, the cation valence $V_{\text{Zr/Y}}$ is the sum of all the bond valences of the nearest neighbors

$$V_{\text{Zr/Y}} = \sum_i s_{\text{Zr/Y},i} \quad (5)$$

We also define the "bond valence energy" $E_{\text{bv}}$ caused by the deviation of the cation valence from the formal charge, $V_{\text{Zr},0} = +4$ and $V_{\text{Y},0} = +3$

$$E_{\text{bv}} = A_{\text{Zr}} \sum (V_{\text{Zr}} - V_{0,\text{Zr}})^{\alpha_{\text{Zr}}} + A_{\text{Y}} \sum (V_{\text{Y}} - V_{0,\text{Y}})^{\alpha_{\text{Y}}} \quad (6)$$

where the sum runs over all the cations within the supercell and $\alpha_{Zr/Y}$ and $A_{Zr/Y}$ are constants. Below, we use $\alpha = 2$ and $A_{Zr} = A_Y = 1$ for simplicity.

We analyzed the 11 structures together with the "ground" state configuration displayed in **Fig. 6**. Because of the variations in bond lengths (in the ground state: Zr-O from 0.20-0.31 nm, Y-O from 0.21-0.27 nm, **Fig. 9a**) and coordination numbers (in the ground state: 7 or 8 for most Zr and Y, 6 for a few Zr), the bond valences of individual Zr and Y vary from +3.4 to +4.0 for Zr and from +3.0 to +3.9 for Y as shown in **Fig. 9b-d**. Despite these variations, however, the average bond valences in **Fig. 9c-d** strongly correlate with the supercell energy obtained by *ab initio* calculations. The ground state, having the lowest energy, has the highest (average) bond valence for Zr—though still less than 4, and the lowest (average) bond valence for Y—though still more than 3. The deviations from formal charges indicate considerable underbonding for Zr and overbonding for Y (even in the "best" structure). Such deviations result in a penalty of a higher $E_{bv}$ shown in **Fig. 9e**. Therefore, although *ab initio* relaxations ensure vanishing forces on all particles, the cation solid solution intermingled with so many $V_O$ cannot fully satisfy all the "ideal" chemical bonding requirements at the local level.

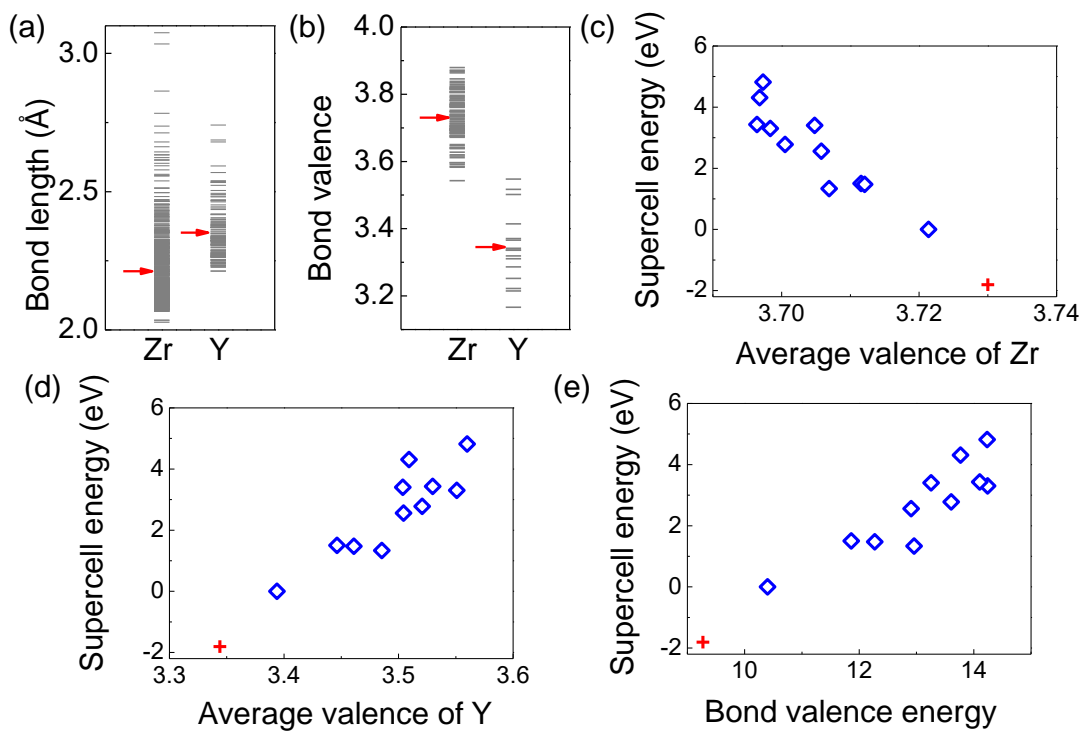

**Figure 9** According to *ab initio* calculations, in the "ground" state the bond lengths (a) and bond valences (b) of both Zr and Y spread widely outside their respective averages indicated by red arrows. Likewise, compared to the same configurations (blue diamonds in (c-e)) in Fig. 5, the ground state (red crosses) has a higher average bond valence of Zr (c), a lower average bond valence of Y (d), and lower bond valence energy (e).

Since overbonding/underbonding indicates deviation from local charge neutrality, it should lead to a higher long-range electrostatic energy for the ionic structure. However, *ab initio* calculations, unlike the empirical potential calculations, do not provide the electrostatic energy as a separate result. So we naively calculated this energy for the *ab initio* relaxed structure by (a) assigning the formal charge to each ion and (b) performing the Ewald sum with the periodic boundary conditions. The results shown in **Fig. 10** do positively correlate with the supercell energy from the *ab initio* calculation, but its magnitude is 3-4 times too large (spanning over ~18 eV in the Ewald sum vs. ~5 eV in the *ab initio* calculations in **Fig. 5**). This problem has already been noted before in **Fig. 5**, which compares the energies of *ab initio* calculations and empirical potential calculations, the latter also include the Ewald sum and use the formal charge (Eq. (1)). In the literature, Bogicevic *et al*. [35] noted the same problem: In their Fig. 10 of Ref. [35], the 2.5 eV/cation energy span in the Ewald sum where formal charge was used is much larger than the 0.5 eV/cation energy span in the *ab initio* calculations. So it appears that the use of the formal charge is likely to overestimate the electrostatic energy.

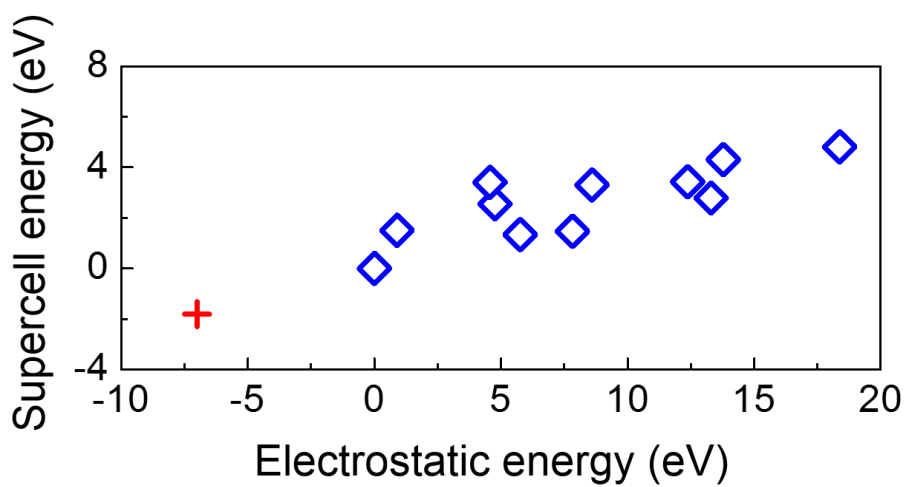

**Figure 10** Electrostatic energy calculated by Ewald sum. Blue diamonds are the same configurations as in **Fig. 5**, and red cross is the "ground" state configuration.

We also calculated the Bader charges on each atom using the charge densities provided by the *ab initio* calculations. [36] (Bader defined the zero-flux surface around an atom by the loci where charge density reaches a local minimum, and the charge enclosed by the surface is the Bader charge of the atom.) The average Bader charge is 0.63 for Zr, 8.81 for Y and 7.66 for O. Relative to the number of electrons used in the PAW potential—4 for Zr, 11 for Y and 6 for O—the average charge of Zr ion is +3.37, Y ion is +2.19, and O ion is -1.66. These numbers will reduce the Ewald sum of the electrostatic energy by about 1.5 times, but an overestimate still remains possibly due to the poor assumption of point charge interactions.

## VI. Discussion

In computational studies, it is not uncommon to first use empirical potential calculations to survey the configurational space, then identify the most promising configurations for further study by *ab initio* calculations [37]. The approach is justified if the calculated energies by the two methods are closely correlated to each other, which is the case here as shown in **Fig. 5**. (A small energy scatter in empirical potential calculations is not necessary as long as the correlation is tight. For example, in **Fig. 5** the ten structures with the lowest energies from empirical potential calculations do span about 3 eV in *ab initio* calculated energies, but they all lie in the lower range of *ab initio* calculated energies in the plot, with four of them having the lowest energies of all the structures studied by *ab initio* calculations. Since the final calculations will be performed using *ab initio* calculations, errors in the empirical potential calculations are unimportant as long as they do not cause any trend reversal thus affecting screening. This is the case in our study.) Obviously, the ease of empirical potential calculations allows their use to explore a much larger sample space. These calculations are especially useful for acquiring certain information such as radial distribution functions and packing preferences from a relatively small sample size: Above a certain size, statistically significant trends with size-invariant ($10^3$-$10^5$ configurations in our case) means and standard deviations are often recognizable as shown in **Fig. 1-3**. On the other hand, statistical sampling is entirely ineffective for locating the low-energy extremes in our 8YSZ study, and in this respect empirical calculations despite their simplicity are of no help. Specifically, when we increased the sample size for empirical potential calculations from $10^3$, to $10^4$, to $10^5$ configurations, the low-energy extremes according to the empirical potential calculations decreased by 0.65 eV and 1.08 eV, respectively, corresponding to about 0.3 eV and 0.5 eV decrement in the *ab initio* calculations if we use the correspondence in **Fig. 5** to convert the energy scales. Extrapolating the latter energy decrement as a function of sample size, we predict a size of ~$10^9$ configurations is needed to achieve the same 1.8 eV decrement as we accomplished in **Fig. 8** by utilizing a packing-rule-based search protocol. Therefore, a search based on appropriate preferences is vastly more efficient than random sampling for locating the lowest energy state in zirconia solid solutions. More broadly, for such problem we would recommend the following approach: (a) Data mining using a modest sized library from empirical potential calculations to establish preferences, (b) preference-directed search again aided by empirical potential calculations, and (c) *ab initio* calculations to verify the outcome of (b).

While the above computational approach may help find a lower energy state, we may have to reconcile the fact that it is practically impossible to find the ground state of 8YSZ. The fundamental reason for this lies in the complexity of the structure: In a rather small 3×3×3 supercell, there are already $N=5.55\times10^{42}$ possible configurations. With such a large $N$, even a survey of $N^{1/2}=2.36\times10^{21}$ configurations is unlikely to encounter the ground state statistically. Moreover, even if we did find the ground state of the 3×3×3 supercell, there is no assurance that it bears any resemblance to the ground state of a macroscopic 8YSZ crystal. Therefore, finding the ground state of a macroscopic 8YSZ and, for that matter, of any macroscopic solid solution, especially a defect-rich one, is truly a daunting challenge.

We now argue that in reality, the ground state is rarely attained in such crystal, the process to approach the ground state is likely to end at some metastable state, and that below certain temperature the crystal may best be considered as a glass frozen at various metastable states some distance away from the ground state. In 8YSZ, the physical reason for the above is obvious. (a) While oxygens can easily migrate to lower the energy of the crystal, cations that diffuse one-trillion-fold slower at below 1,000 °C cannot relocate themselves fast or far enough to significantly lower the system energy. (b) YSZ contains two cations of different charge and size, which demand contradicting packing preferences and bond valence requirements. (c) The constraint of composition exacerbates the difficulty in satisfying (b), as evidenced by the large variations and far-from-ideal (i) coordination numbers in **Fig. 3** and **6**, and (ii) bond lengths/valences in **Fig. 9a-b**, despite the tendency for the coordination numbers, bond lengths and bond valences to well behave *on average*. The notion of glassy structure and glass transition is supported by the following experimental observations: Doped zirconia systems have non-Arrhenius dependence of ionic conductivity with an increased apparent activation energy at <500°C, with such data better fit by the Vogel-Fulcher-Tammann equation of glassy kinetics than by a combination of Arrhenius equations of single or multiple diffusion processes/paths [38, 39].

Although our study is focused on the energetics of 8YSZ, our computational experience in collecting data for **Fig. 7-8** did provide insight into how 8YSZ may kinetically lower its energy. To recap the experience: We first started with a randomly selected configuration of the cation sublattice, then filled $V_O$ at the most preferred sites to maximize $n_{Y\text{-}V_O\text{-}2NN}$, which is strongly correlated with a lower supercell energy. As $n_{Y\text{-}V_O\text{-}2NN}$ increases (i.e., energy decreases), we found for some starting cation superlattices it became difficult to further decrease the energy by moving $V_O$ alone; some rearrangement of the cation sublattices must be undertaken for this to happen. Eventually, we reached some maximum $n_{Y\text{-}(V_O\text{-}2NN)}$ (*aka* minimum energy) because we could not find any suitable cation sublattice to lower the energy further despite trying a reasonably large number of times. Yet the $n_{Y\text{-}(V_O\text{-}2NN)}$ (52) at this point is still far smaller than the theoretical maximum $n_{Y\text{-}(V_O\text{-}2NN)}$ (96) for the supercell. While the above was our experience with $n_{Y\text{-}(V_O\text{-}2NN)}$, which pertains to 2NN, we had an entirely different experience with maximizing $n_{Zr\text{-}(V_O\text{-}1NN)}$, which is the number of the 1NN pairs. We found the latter task rather easy and we had no difficulty in reaching the theoretical maximum $n_{Zr\text{-}(V_O\text{-}1NN)}$ (32). In experimental terms, these computational observations may be interpreted as follows. The relaxation processes in a YSZ "glass" ranges from very fast ones (V<sub>O</sub> diffusion), to intermediate ones (short-range rearrangement of cation sublattice) to the slowest ones (long-range rearrangement of cation sublattice). While all relaxations can lower the energy, very fast and intermediate processes can only reach some metastable states; to reach metastable states of a much lower energy, cooperative long-range relaxations are mandatory. In general, the relaxation time requiring cation diffusion increases rapidly with the length scale of diffusion and the number of ions involved. So as the length scale becomes macroscopic, the relaxation time diverges below certain temperature (i.e., the glass temperature). This probably happens to YSZ when the temperature falls below 1,000°C. When it happens, even with long annealing the crystal can only sample some nearly degenerate metastable states but never the true ground state. This is the essence of glass transition in YSZ.

Having associated our general observations in this small-supercell study of YSZ to the phenomenon of glass transition, we wish to emphasize that the strong crystal chemistry preferences that dictate the packing rules will accentuate the tendency toward such transition. This is because such rules impose long-range correlations persisting to the 5NN, which means that the preference commanded by a center cation in a 3×3×3 supercell can dictate the atomic arrangement at the outer boundary of the supercell. This is not an artifact that arises from the small size of the supercell, since the same trend was observed as we repeated the calculations in the 4×4×4 and 5×5×5 supercells. Inasmuch as a macroscopic YSZ sample may be regarded as a collection of many small supercells of different cation configurations, we can appreciate the frustration at the supercell-supercell boundaries, where the conflicting packing preferences of neighboring supercells may be impossible to reconcile. This, naturally, will lead to glass transition.

Lastly, we briefly remark on our observations of crystal chemistry preferences and bond valence. In zirconia solid solutions, whether $V_O$ prefers to be the 1NN or 2NN of Zr and dopant cations has been customarily explained by the relative size and charge of these cations [6, 35]. Our empirical-potential and *ab initio* calculations have confirmed such preferences in 8YSZ, but the remarkable discovery of additional preferences that extend all the way out to at least the 5NN in an oscillation pattern is entirely unexpected. Moreover, such long-range oscillations, which must come from similarly long-range effects including elastic and electrostatic interactions, are especially remarkable in view of the complete washout of the ionic size effect on the radial distribution functions at distances beyond the 1NN environment. Interestingly, atomic scale oscillations of Zr/Y/O concentrations have also been recently observed, in two dimensions, near $\Sigma 3$ and $\Sigma 5$ symmetric tilt grain boundary in experimental [40] and simulation studies [41]. These two- and three-dimensional oscillations may share the same mechanistic origin, of which we suggest two possibilities. (a) A finite ion size effect in analogy to the minimum electron wavelength (the Fermi wavelength) effect. For charge screening, the latter is known to lead to the Friedel oscillations as opposed to the monotonic, basically exponential Debye-Hueckel decay predicted by the continuum theory. (b) A multiple-like ordering of $V_O$ and cations, which optimizes the interplay between elastic and electrostatic interactions. Unfortunately, the present study on YSZ that contains only one dopant (Y) cannot probe the effects of different dopant charge, size and elastic modulus. So it is unable to differentiate these effects and the two mechanisms. On the other hand, the long-range nature of crystal chemistry preferences strongly implies any short-range parameter such as bond valence is insufficient by itself for the determination of the lowest-energy configuration or even the local structure. Indeed, the large variation of bond valences (meaning the broad violations of the bond valence rules) in the structure of the "ground" state (**Fig. 9c-d**) apparently has no immediate consequence on the forces on ions/electrons or the total energy of the material.

## VII.    Conclusions

(1) YSZ as a defect-rich concentrated solid solution has numerous metastable configurations with energies well above that of the ground state. Such metastable states are unlikely to realize in real YSZ, so computation of their properties has no practical meaning. Meanwhile, being an extreme-energy state, the ground state cannot be accessed by statistical sampling.

(2) The energies of randomly sampled configurations follow a robust set of correlations to their local structures. The following packing rules are preferred: $V_O$ should be the 1NN, 3NN and 5NN of Zr and the 2NN and 4NN of Y, with the formation of 2NN $V_O$-Y pairs being the strongest preference. Conventional local probes (such as EXAFS) cannot detect these features because the average radial distribution functions do not manifest any size effect beyond the first cation-O shell.

(3) An intelligent sampling method incorporating the packing rules can efficiently search for the lowest energy configurations to find a *de-facto* "ground" state; this state will be used for cation diffusion calculations in the companion paper.

(4) Practical applications of YSZ almost invariably involve a glassy state. It can partially relax with rapid oxygen diffusion, but it cannot fully relax because the extremely slow cation diffusion is frustratingly hampered by structural complexity and by long-ranged crystal chemistry coupling.

**Acknowledgements**

This work was supported by the Department of Energy (BES grant no. DEFG02-11ER46814) and used the facilities (LRSM) supported by the U.S. National Science Foundation (grant no. DMR-1120901). JL acknowledges support by NSF DMR-1410636.